# Learning-based Array Configuration-Independent Binaural Audio Telepresence with Scalable Signal Enhancement and Ambience Preservation

Yicheng Hsu, and Mingsian R. Bai, *Senior Member, IEEE*

*Abstract*—Audio Telepresence (AT) aims to create an immersive experience of the audio scene at the far end for the user(s) at the near end. The application of AT could encompass scenarios with varying degrees of emphasis on signal enhancement and ambience preservation. It is desirable for an AT system to be scalable between these two extremes. To this end, we propose an array-based Binaural AT (BAT) system using the DeepFilterNet as the backbone to convert the array microphone signals into the Head-Related Transfer Function (HRTF)-filtered signals, with a tunable weighting between signal enhancement and ambience preservation. An array configuration-independent Spatial COherence REpresentation (SCORE) feature is proposed for the model training so that the network remains robust to different array geometries and sensor counts. magnitude-weighted Interaural Phase Difference error (mw-IPDe), magnitude-weighted Interaural Level Difference error (mw-ILDe), and modified Scale-Invariant Signal-to-Distortion Ratio (mSI-SDR) are defined as performance metrics for objective evaluation. Subjective listening tests were also performed to validate the proposed BAT system. The results have shown that the proposed BAT system can achieve superior telepresence performance with the desired balance between signal enhancement and ambience preservation, even when the array configurations are unseen in the training phase.

*Index Terms*—binaural reproduction, audio telepresence, spatial feature, microphone array, deep learning

## I. INTRODUCTION

VIRTUAL reality is a fast-growing field with applications not only in teleconferencing and broadcasting, but also in virtual concerts, filming, ball game recording, and virtual guided tours, to name a few [1]-[4]. As a form of audio virtual reality, Audio Telepresence (AT) aims to transport the acoustic scene at the far end to the user(s) at the near end in such a way that the spatial impression of sources or ambient sounds at the far end is preserved and reproduced at the near end. The objective of AT is to present immersive and interactive audio between locations with near life-like audio quality. At the near-end, either headphones [5]-[7] or loudspeaker arrays [8]-[10] can be used as the rendering device for AT, where the idea behind the latter rendering approach is similar to the sound field reconstruction techniques such as Ambisonics [11, 12] and Wave Field Synthesis (WFS) [13, 14]. In this study, we will focus on the Binaural Audio Telepresence (BAT).

Ambience preservation and the signal enhancement are two primary concerns of AT. However, depending on the application scenarios, we may place different emphasis on these two aspects. For in-car entertainment, we want to experience the full ambience as if we were at the far end, while for teleconferencing, we want to hear the fully enhanced far-end conversation with high quality and intelligibility. Specifically, traditional speech enhancement seeks to suppress noise, interference and reverberation. This is not to be confused with source separation problems, where the enhanced signals are required to be further separated into independent source signals. Ambience preservation seeks to preserve the entire acoustic scene, regardless of the source signal or the ambient sound. A typical BAT process may involve a spherical microphone array for sound capture [15, 16], followed by Head related transfer function (HRTF) filtering [17]-[20]. Signal enhancement may be required to suppress interference, noise, and reverberation. [21]-[23].

There are two design principles for a BAT system that can be found in the literature. A three-stage Localization-Beamforming-HRTF Filtering (LBH) approach was suggested [24]. This approach can provide excellent binaural output with accurate directional perception at the expense of audio ambience. Another approach that generally works better for ambience preservation is based on the Model-Matching Principle (MMP). For example, beamforming-based binaural reproduction [25] and Multichannel Inverse Filtering (MIF) [26] were investigated on the basis of MMP. In addition, learning-based Multichannel Deep Filtering (MDF) [26] has been proposed to preserve the spatial impression of the target source and ambient sounds, while suppressing noise and reverberation. Although signal enhancement is considered an important element of AT in many application scenarios such as virtual meetings, ambience preservation has been overlooked, but can be equally important in some other scenarios such as in-car entertainment. A scalable AT system that can seamlessly migrate between these two extremes would be highly desirable. To this end, an Array

This work was supported by the National Science and Technology Council (NSTC), Taiwan, under the project number 110-2221-E-007-027-MY3. *(Corresponding author: Mingsian R. Bai).*

Yicheng Hsu is with the Department of Power Mechanical Engineering, National Tsing Hua University, Hsinchu, Taiwan (e-mail: shane.ychsu@gmail.com).

Mingsian R. Bai is with the Department of Power Mechanical Engineering and Electrical Engineering, National Tsing Hua University, Hsinchu, Taiwan (e-mail: msbai@pme.nthu.edu.tw).

Configuration-Independent Scalable Binaural AT (ACIS-BAT) system is proposed in this study.

ACIS-BAT can be considered as a hybrid system that combines a Digital Signal Processing (DSP)-based feature extractor and a learning-based AT deep neural network that can accommodate different array configurations of geometry and channel count. As a core element of ACIS-BAT, an array configuration-independent Spatial COherence REpresentation (SCORE) module is developed to extract spatial features robust to array configuration variations. With the extracted spatial features, we propose a Binaural Rendering network (BRnet) based on a DeepFilterNet [27] to transform the microphone signals into binaural signals. The scalability between signal enhancement and ambience preservation is achieved through a tunable parameter in BRnet injected into a Feature-wise Linear Modulation (FiLM) layer [28].

Experiments are performed to validate the proposed binaural AT system against to several baselines, including DSP-based LBH and MIF methods and the learning-based MDF method. To assess the robustness of the proposed system with respect to different array configurations, unseen array geometries and varying numbers of microphones are used in the testing phase of our learning-based method. Since the Interaural Phase Difference (IPD) and the Interaural Level Difference (ILD) are two important spatial cues for localization [29, 30], we introduce the magnitude-weighted IPD error (mw-IPDe) and the magnitude-weighted ILD error (mw-ILDe) as two objective metrics to evaluate the spatial attribute of the binaural output signals. In addition, the modified Scale-Invariant Signal-to-Distortion Ratio (mSI-SDR) is defined to evaluate the reproduction quality and enhancement performance. Apart from the objective evaluation, listening tests are performed based on subjective indices, including sense of direction, ambience preservation, background noise reduction, sensor noise reduction, dereverberation, and overall quality for signal enhancement and ambience preservation.

The main contributions of this paper can be summarized as follows:

- We present an AT paradigm under which a novel array-based binaural rendering system that is robust to unseen array configurations is proposed.
- We use a hybrid approach to implement the proposed system. DSP extracts spatial features, whereas DNN converts the array signals into the binaural outputs.
- An array configuration-independent Spatial COherence REpresentation (SCORE) is proposed to serve as an effective spatial feature for the rendering network.
- The notion of scalability between signal enhancement and ambience preservation is introduced to accommodate more AT application scenarios.
- Three objective metrics are suggested to evaluate the performance of BAT systems.

The remainder of this paper is organized as follows. Section II outlines the signal model and the problem formulation. Section III introduces the proposed array configuration-independent binaural rendering system. In Sec. IV, we present the experimental results of the proposed method compared to several baselines. The conclusions are given in Sec. V.

## II. SIGNAL MODEL AND PROBLEM FORMULATION

Consider a scenario where a target speaker and ambient sound signals are picked up by $M$ microphones in a reverberant room. The received signal at the $m$th microphone can be written in the Short-Time Fourier Transform (STFT) domain as

$$X^m(l,f) = H_t^m(f)S_t(l,f) + \sum_{j=1}^{J} H_j^m(f)S_a(l,f) + V^m(l,f), \quad (1)$$

where $l$ and $f$ denote the time frame index and frequency bin index, respectively, $H_t^m(f)$ denotes the Acoustic Transfer Function (ATF) between the $m$th microphone and the target speaker, $H_j^m(f)$ denotes the ATF between the $m$th microphone and the source in the $j$th direction, $S_t(l,f)$ denotes the signal of the target speaker, $S_a(l,f)$ denotes the ambient source signal, and $V^m(l,f)$ denotes the additive sensor noise. The ambient source signal incident from $J$ directions equally spaced on the horizontal plane are used to simulate the ambient sound ($J = 72$ in this study).

The goal of the BAT system is to convert the microphone array signals into the binaural signals with a prescribed balance between full ambience and full enhancement. The target binaural output signals are given as

$$Y_L(l,f) = H_{s,L}^{hrtf}(f)H_{t\_clean}^m(f)S_t(l,f) + \alpha \sum_{j=1}^{J} H_{j,L}^{hrtf}(f)H_j^1(f)S_a(l,f),$$

$$Y_R(l,f) = H_{s,R}^{hrtf}(f)H_{t\_clean}^m(f)S_t(l,f) + \alpha \sum_{j=1}^{J} H_{j,R}^{hrtf}(f)H_j^1(f)S_a(l,f), \quad (2)$$

where $H_{t\_clean}^m(f)$ denotes the direct and early part of the ATF $H_t^m(f)$, $H_{s,L}^{hrtf}(f)$ and $H_{s,R}^{hrtf}(f)$ denote the HRTFs from the target speaker to the left and right ears, $H_{j,L}^{hrtf}(f)$ and $H_{j,R}^{hrtf}(f)$ denote the HRTFs from the $j$th ambient source to the left and right ears, and $\alpha \in [0,1]$ is a scaling factor to strike a balance between full enhancement ($\alpha = 0$), and full ambience ($\alpha = 1$).

## III. PROPOSED SYSTEM

In this section, an array-based BAT system, ACIS-BAT, using a DNN as the backbone to convert the array microphone signals into the HRTF-filtered signals, with a tunable weighting between signal enhancement and ambience preservation is proposed. The block diagram of the ACIS-BAT system is shown in Fig. 1, where the Spatial COherence REpresentation (SCORE) module provides spatial features to the subsequent Binaural Rendering network (BRnet), enabling array configuration-independent applications. Note that a tunable parameter $\alpha$ defined in Eq. (2) can be used to weight between the signal enhancement and ambience preservation modes according to the application scenario. The DSP-based SCORE module is described next.

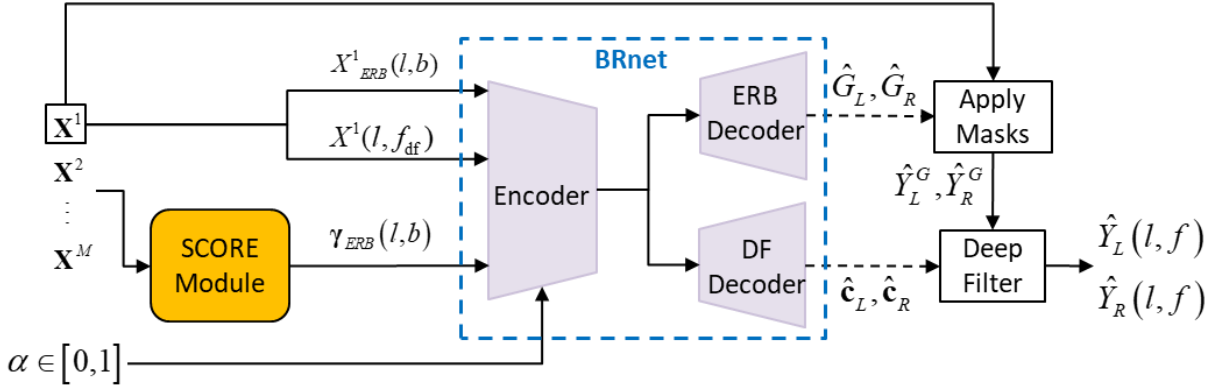

**Fig. 1.** The block diagram of the ACIS-BAT system. The learnable modules are highlighted by the dashed blue box, while the rest are non-learnable modules.

### A. The SCORE Module

For each TF bin, the short-term relative transfer function (RTF) between the $m$th microphone and reference microphone 1 can be estimated by averaging $(R + 1)$ frames:

$$\tilde{R}^m(l,f) \equiv \frac{\sum_{n=l-R/2}^{l+R/2} X^m(l,f) X^{1*}(l,f)}{\sum_{n=l-R/2}^{l+R/2} X^1(l,f) X^{1*}(l,f)}, \quad (3)$$

where * denotes the complex conjugate operation. As a key step, a "whitened" feature vector $\mathbf{r}(l,f) \in \mathbb{R}^{(M-1)\times 1}$ associated with each TF bin is computed as follows:

$$\mathbf{r}(l,f) = \left[ \frac{\tilde{R}^2(l,f)}{|\tilde{R}^2(l,f)|}, \cdots, \frac{\tilde{R}^M(l,f)}{|\tilde{R}^M(l,f)|} \right]^T, \quad (4)$$

where $|\bullet|$ denotes the complex modulus.

Assuming $D$ candidate zones in azimuthal angles, the corresponding free-field plane-wave RTF vector of the $j$th zone can be written as

$$\mathbf{a}_j(f) = \left[ e^{-j\mathbf{k}_j \cdot (\mathbf{p}_2 - \mathbf{p}_1)} \quad \cdots \quad e^{-j\mathbf{k}_j \cdot (\mathbf{p}_M - \mathbf{p}_1)} \right] \in \mathbb{C}^{(M-1)\times 1}, \quad (5)$$

where $\mathbf{p}_m$ is the position vector of the $m$th microphone, $\mathbf{k}_j = -k\mathbf{\kappa}_j = -(\omega/c)\mathbf{\kappa}_j = -(2\pi f/c)\mathbf{\kappa}_j$ is the wave vector of the $j$th plane-wave component, $\mathbf{\kappa}_j$ denotes a unit vector pointing to the look direction, $k$ denotes the wave number, and $c$ is the sound speed. To exploit the spatial information conveyed by the whitened RTF, we define the SCORE feature vector as

$$\boldsymbol{\gamma}(l,f) \approx \frac{1}{M-1} \mathrm{Re}\left\{ \mathbf{A}^H(f)\mathbf{r}(l,f) \right\} \in \mathbb{R}^{J\times 1}, \quad (6)$$

where $\mathbf{A}(f) = [\mathbf{a}_1(f) \quad \cdots \quad \mathbf{a}_j(f)] \in \mathbb{C}^{(M-1)\times J}$ and $\mathrm{Re}\{\cdot\}$ denotes the element-wise real-part operator. The Euclidean angle [31] rather than the Hermitian angle is adopted in SCORE definition for sign sensitivity. Note that the number of microphones affects the dimension of the feature vector in Eqs. (4) and (5), not the dimension of the SCORE vector $\boldsymbol{\gamma}(l,f)$. In addition, only the position vector $\mathbf{p}_m$ in Eq. (5)

should be modified for different array geometries. Therefore, the model trained with SCORE is very robust to variations in array configuration.

The STFT-domain SCORE feature is computationally expensive. A preferred method is to compress the SCORE feature in the Equivalent Rectangular Bandwidth (ERB) scale [32]:

$$\boldsymbol{\gamma}_{ERB}(l,b) = \frac{1}{\pi_b} \sum_{f \in \{f_{b1},\ldots,f_{bF_b}\}} w_b(f)\boldsymbol{\gamma}(l,f), \ b \in \{0,1,\ldots,B\}, \quad (7)$$

where $B$ is the total number of bands ($B = 48$ in this paper), $w_b(f)$ and $F_b$ are the weight and the number of the frequency bins for the $b$th band, and $\pi_b$ is a normalization factor defined as

$$\pi_b = \sum_f^{F_b} w_b(f), \quad (8)$$

### B. The BRnet

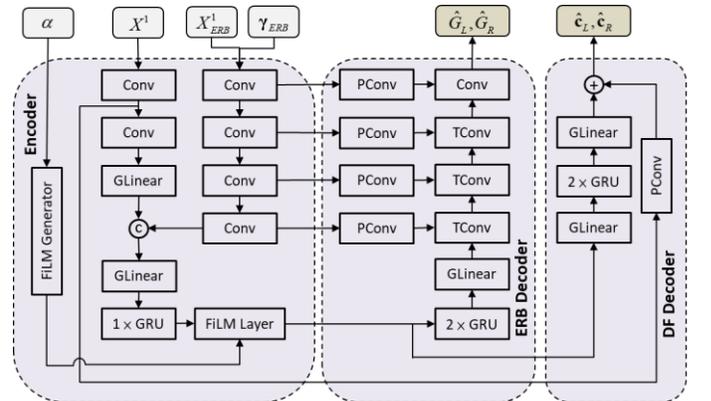

**Fig. 2.** The architecture of BRnet. The Encoder extracts the spatial and spectral information for the subsequent ERB Decoder and DF Decoder to generate the masks and deep filter coefficients for the reference microphone.

For efficient DNN operation, BRnet is based on DeepFilterNet [27], which performs a two-stage denoising process with reference to human auditory properties. The BRnet, which consists of three modules, the Encoder, the ERB Decoder and the Deep Filter (DF) Decoder, is shown in Fig. 2.

The network operation can be formulated as follows:

$$\mathbf{e}(l) = \mathcal{F}_{enc}\left(Y_{ERB}^1(l,b), Y_{DF}^1(l,f_{DF}), \gamma_{ERB}(l,b)\right), \quad (9)$$

where an encoder $\mathcal{F}_{enc}$ encodes the ERB feature and the STFT feature of the reference microphone signal and the ERB-scaled SCORE feature $\gamma_{ERB}$ into an embedding $\mathbf{e}$. Instead of concatenating the factor $\alpha$ with the embedding, we use a FiLM layer [28] to apply a feature-wise affine transformation, including scaling and shifting operations, to the embedding. This affine transformation is useful for learning conditional representations. The FiLM generator accepts the tunable parameter $\alpha$ as the input and produces a scaling vector $\beta(\alpha)$ and a shifting vector $\delta(\alpha)$, with the same dimension as $\mathbf{e}$. The output of the FiLM can be expressed as

$$\text{FiLM}(\mathbf{e}(l)) = \beta(\alpha) \cdot \mathbf{e}(l) + \delta(\alpha), \quad (10)$$

which is then first passed to the ERB Decoder to predict two real-valued ratio masks ($\hat{G}_L$ and $\hat{G}_R$) for processing the noisy spectrum of $Y^1$. It follows that the spectra, $Y_G^L$ and $Y_G^R$, for the left and right ears can be estimated as

$$G_{erb}^L(l,b), G_{erb}^R(l,b) = \mathcal{F}_{erb\_dec}(\mathbf{e}(l))$$
$$\hat{G}^L(l,f) = \text{ERB}^{-1}\left(G_{erb}^L(l,b)\right), \quad \hat{G}^R(l,f) = \text{ERB}^{-1}\left(G_{erb}^R(l,b)\right) \quad (11)$$
$$\hat{Y}_G^L(l,f) = Y^1(l,f) \cdot \hat{G}^L(l,f), \quad \hat{Y}_G^R(l,f) = Y^1(l,f) \cdot \hat{G}^R(l,f)$$

where $\text{ERB}^{-1}$ denotes the inverse ERB transform. In the filtering stage, the DF Decoder predicts two DF coefficients of order $N$ to obtain the enhanced spectrums of the left and right ears ($\hat{Y}_G^L$ and $\hat{Y}_G^R$) using linear filters:

$$C_L^N(l,i,f_{DF}), C_R^N(l,i,f_{DF}) = \mathcal{F}_{df\_dec}(\mathbf{e}(l))$$
$$\hat{Y}_L(l,f') = \sum_{i=0}^{N} C_L(l,i,f') \hat{Y}_G^L(l-i+q,f'), \quad (12)$$
$$\hat{Y}_R(l,f') = \sum_{i=0}^{N} C_R(l,i,f') \hat{Y}_G^R(l-i+q,f'),$$

where $q$ is the look-ahead of the deep filter. In contrast to [27], the proposed BRnet operates at 16 kHz and uses an ERB filterbank with 32 bands. The deep filter with $N = 5$ is applied only to the lowest 160 frequency bins which covers a 5 kHz bandwidth. The number of input channel of the ERB encoder depends on the number of the input feature channels ($1+J$). To generate the ERB masks for the left and right ears, the number of output channels of the ERB decoder is set to 2. Similarly, two DF coefficients are predicted and applied to the left ear and the right ear to restore the phase information. The remaining parameter settings are the same as in [27].

### C. Training Procedure and Loss Function

In this study, the signal frames were prepared with a length of 32 ms and a step of 8 ms, and a 512-point FFT was used. The optimizer used for training was Adam with a learning rate of 0.001. A gradient norm clipping of 3 was used. The learning rate was halved if there was no improvement in the loss of the validation set for three consecutive epochs. The complex compressed mean-square error [33] was adopted as the loss function, consisting of the magnitude-only and phase-sensitive terms.

$$\mathcal{L} = (1-\lambda) \sum_{l,f} \left\| |\mathbf{Y}|^c - |\hat{\mathbf{Y}}|^c \right\|_F^2 + \lambda \sum_{l,f} \left\| |\mathbf{Y}|^c e^{j\angle\mathbf{Y}} - |\hat{\mathbf{Y}}|^c e^{j\angle\hat{\mathbf{Y}}} \right\|_F^2, \quad (13)$$

where $\mathbf{Y} = \begin{bmatrix} \mathbf{Y}_R & \mathbf{Y}_L \end{bmatrix}$ represents the target binaural signal defined in Eq. (2), $\hat{\mathbf{Y}} = \begin{bmatrix} \hat{\mathbf{Y}}_R & \hat{\mathbf{Y}}_L \end{bmatrix}$ represents the estimated binaural signal defined in Eq. (12), $c = 0.3$ is a compression factor, and $\lambda = 0.2$ is a weighting factor. For the sake of brevity, the frequency and time indices are omitted. Magnitude compression and angle extraction are element-wise operations.

### D. Training the Model for Both SE and AP Tasks

There is a practical need for a scalable AT system that can handle both the SE and the AP tasks or even the conditions between these extremes. However, integrating these two extremes into a single model is challenging. To achieve this, we propose a simple but effective training paradigm to ensure satisfactory performance between signal enhancement and ambience preservation conditions. During each training epoch, the factor $\alpha$ in Fig. 2 and Eq. (2) is randomly sampled from $\{0.0, 0.3, 0.5, 0.7, 1.0\}$. Taking advantage of supervised learning, the model can learn to adjust the value of the ambient component by passing the embedding vector through the FiLM layer based on the sampled factor $\alpha$. The projection of the embedding vector can be visualized in Fig. 3. This allows the user to adjust the tunable factor $\alpha$ to meet different needs without changing the model during testing.

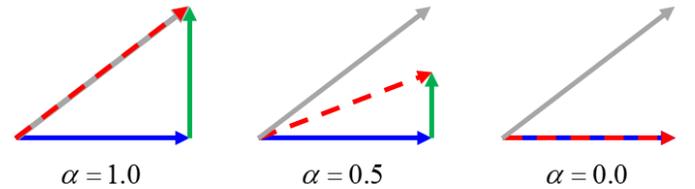

$\alpha = 1.0$     $\alpha = 0.5$     $\alpha = 0.0$

**Fig. 3.** Visualization of embedding vector projection according to different factors $\alpha$. The vectors corresponding to the target, ambient, and noisy signals are represented by the blue, green, and gray lines, respectively. The red dashed line represents the vector projected through the FiLM layer.

## IV. EXPERIMENTAL STUDY

In this section, extensive experiments were conducted to validate the proposed ACIS-BAT system. To investigate the robustness of the spatial feature SCORE, the Modal Assurance Criterion (MAC) is used for different array configurations. Three objective metrics are defined to evaluate its rendering and enhancement performance against several baselines. The scalability of the proposed system is also investigated with respect to the tunable parameter $\alpha$. In addition, subjective listening tests were conducted with various signal enhancement and ambience preservation measures.

## A. Data Preparation

In this study, clean speech signals consisting of utterances form 921 and 40 speakers were selected from the *train-clean-360* and *test-clean* subsets of the LibriSpeech corpus [34] for training and testing. To simulate the ambient sounds, music signals from the Free Music Archive [35] were used. The experiment was performed at a sample rate of 16 kHz. A five-element uniform circular array was used for training, with one microphone in the center and the other four evenly distributed around a 4 cm radius circle. The Room Impulse Responses (RIRs) were simulated by using the image-source method [36] with reverberation time (T60) = 0.2, 0.4, and 0.6 s. The HRTFs were selected from the SADIE II database provided by the University of York [37]. The data preparation procedure for training and validation is shown in Fig. 4. To generate the microphone signals, the speech signal is convolved with the RIR in the target direction, while the music signal is convolved with the RIRs in all directions to simulate ambient sounds. We also convolved not only the RIRs but also the HTFRs to generate the binaural target speech and ambient sounds. In addition, the late reverberation of the RIR for the target speech signal is deliberately attenuated for natural sounding, as in [38].

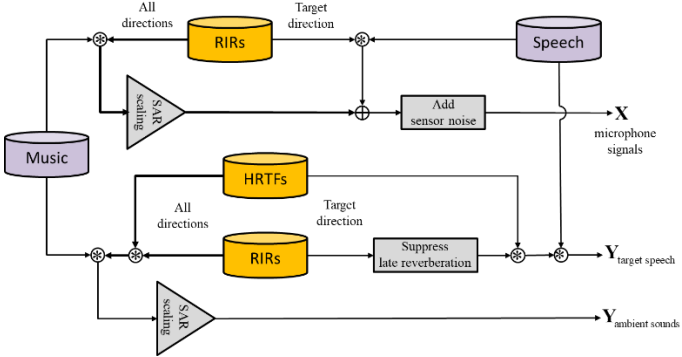

**Fig. 4.** Data preparation of the noisy mixture and target signals for the network training.

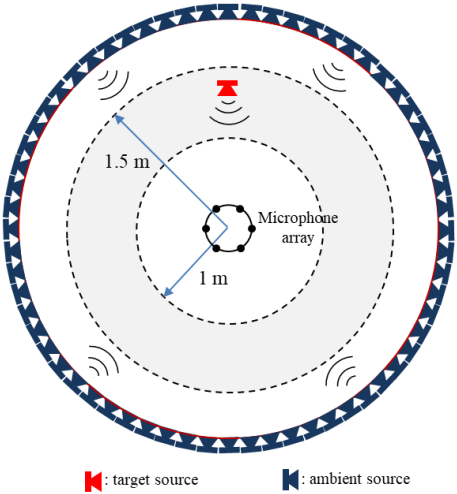

**: target source   **: ambient source

**Fig. 5.** Experimental setup for the training and validation.

The experimental setup for the training and validation is shown in Fig. 5, where music signals played in 72 equally spaced directions simulate the ambient sounds and the target speaker is randomly positioned in the ring sector bounded by radius = 1.0 m and 1.5 m. A total of 50,000 and 5,000 samples are used for training and validation, respectively. Noisy audio signals, edited in five-second clips, are prepared by mixing a target male or female speech signal with the ambient sounds at Signal-to-Ambient Ratio (SAR) = 0, 5, 10, and 15 dB. In addition, sensor noise is added with Signal-to-Noise Ratio (SNR) = 20, 25, and 30 dB. In this study, simulated Gaussian white noise was used to simulate the sensor noise.

In the testing phase, we generate 400 samples of test data to evaluate the proposed ACIS-BAT system. We consider a scenario where a target speaker source moves in a circle with SAR = 10 dB and SNR = 25 dB, as illustrated in Fig. 6. Reverberation is simulated for T60 = 340 and 460 ms. In addition, four different array configurations are used to assess the robustness of BAT to unseen array geometries and unknown numbers of microphones, as shown in Fig. 7. The first array configuration (G1) is included in the training set, while the remaining array configurations (G2, G3, and G4) are "unseen" to the trained model.

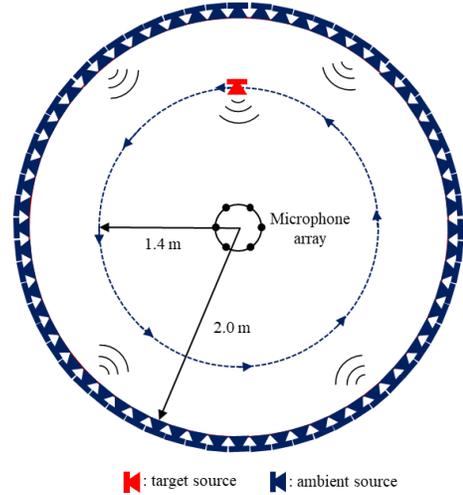

**: target source   **: ambient source

**Fig. 6.** Experimental setup. The microphone array, the moving target source, and the ambient sources are indicated.

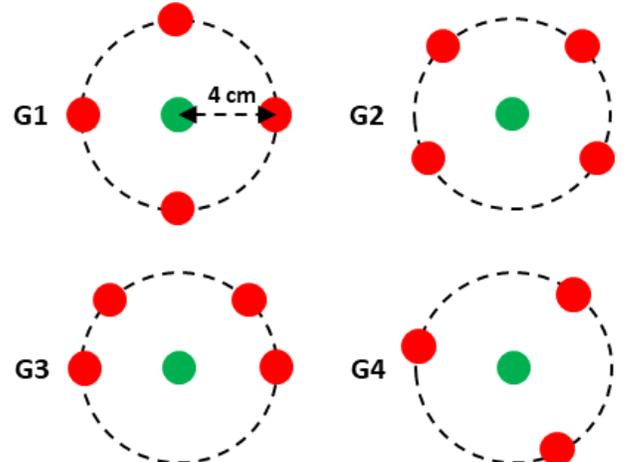

**Fig. 7.** Microphone array settings used in the experiments to investigate the effects of array geometry and the number of microphones. The green dots indicate the reference microphones.

*B. Implementation details*

Three baselines are used to benchmark the proposed ACIS-BAT system. The first baseline is the LBH approach which performs source localization, signal extraction, and HRTF filtering in tandem. The first two stages involved the use of the steered-response power phase transform [39] and the Minimum Power Distortionless Response (MPDR) beamformer [40], with free-field steering vectors used in both stages. HRTFs are determined by the localization result. The second baseline is the MIF approach [26] based on the mode-matching principle, where HRTFs corresponding to 72 ambient source directions equally spaced on the horizontal plane are selected in the desired model. The room impulse responses for MIF were simulated using the image-source method [36]. The regularization parameter used in MIF is set to 0.0001. Unlike the first two DSP-based baselines, the third baseline, MDF, is a learning-based method [26]. The MDF network (MDFnet) can only be optimized for either signal enhancement or ambience preservation, not both. Therefore, two MDFnets were trained with $\alpha$ set to 0 for Signal Enhancement (MDF-SE) and trained with $\alpha$ set to 1 for Ambience Preservation (MDF-AP). Table I summarizes the required parameter size and floating point operations (FLOPs) for MDFnet and the proposed BRnet. Note that BRnet requires only half the parameter size and FLOPs compared to the previously published MDFnet.

TABLE I
NUMBER OF PARAMETERS AND FLOPS FOR MDFNET AND THE PROPOSED BRNER

| Model | Params [M] | FLOPs [M] |
|-------|-----------|-----------|
| MDFnet | 6.36 | 4.00 |
| BRnet | **2.50** | **1.91** |

*C. Spatial Feature Robustness*

In this section, we will evaluate the robustness of the spatial feature provided by the MDF and the proposed methods when applied to unseen array geometries and different numbers of microphones. For the MDF method, we extract the InterChannel Phase Difference (ICPD) as the spatial feature. We vectorize the ICPD feature as

$$\psi = \begin{bmatrix} \boldsymbol{\varphi}^T(l_1) & \boldsymbol{\varphi}^T(l_2) & \cdots & \boldsymbol{\varphi}^T(l_L) \end{bmatrix}^T \in \mathbb{R}^{(M-1)KL \times 1}, \quad \text{where}$$

$$\boldsymbol{\varphi}(l) = \begin{bmatrix} \boldsymbol{\varphi}^T(l, f_1) & \boldsymbol{\varphi}^T(l, f_2) & \cdots & \boldsymbol{\varphi}^T(l, f_K) \end{bmatrix}^T \in \mathbb{R}^{(M-1)K \times 1} \text{ and}$$

$$\boldsymbol{\varphi}(l, f) = \begin{bmatrix} IPD_2(l, f) & IPD_3(l, f) & \cdots & IPD_M(l, f) \end{bmatrix}^T \in \mathbb{R}^{(M-1) \times 1}$$

. For the proposed method, we define the SCORE and ERB-scaled SCORE as the spatial features. SCORE is vectorized as

$$\psi = \begin{bmatrix} \boldsymbol{\gamma}^T(l_1) & \boldsymbol{\gamma}^T(l_2) & \cdots & \boldsymbol{\gamma}^T(l_L) \end{bmatrix}^T \in \mathbb{R}^{JKL \times 1}, \quad \text{where}$$

$$\boldsymbol{\gamma}(l_1) = \begin{bmatrix} \boldsymbol{\gamma}^T(l_1, f_1) & \boldsymbol{\gamma}^T(l_1, f_2) & \cdots & \boldsymbol{\gamma}^T(l_1, f_K) \end{bmatrix}^T \in \mathbb{R}^{JK \times 1}. \quad \text{The}$$

ERB-scaled SCORE is vectorized as

$$\psi = \begin{bmatrix} \boldsymbol{\gamma}_{ERB}^T(l_1) & \boldsymbol{\gamma}_{ERB}^T(l_2) & \cdots & \boldsymbol{\gamma}_{ERB}^T(l_L) \end{bmatrix}^T \in \mathbb{R}^{JBL \times 1}, \quad \text{where}$$

$$\boldsymbol{\gamma}_{ERB}^T(l_1) = \begin{bmatrix} \boldsymbol{\gamma}_{ERB}^T(l_1, f_{b1}) & \boldsymbol{\gamma}_{ERB}^T(l_1, f_{b2}) & \cdots & \boldsymbol{\gamma}_{ERB}^T(l_1, f_{bF_s}) \end{bmatrix} \in \mathbb{R}^{JB \times 1}$$

. Let $\psi$ and $\psi'$ be the feature vectors associated with two array

configurations. We can compute Modal Assurance Criterion (MAC) as follows

$$MAC(\psi, \psi') = \frac{(\psi^T \psi')^2}{\psi^T \psi \psi'^T \psi'}, \quad (14)$$

To evaluate the robustness of ICPD, SCORE, and ERB-scaled SCORE, the MAC values are computed for different array configurations (G1, G2, G3, and G4), as are shown in Fig. 7, using the test datasets. Tables II, III, and IV summarize the MAC values obtained using ICPD, SCORE, and ERB-scaled SCORE. Note that the off-diagonal MAC values in Table II are much smaller than one, indicating a low degree of correlation between the ICPD values for different array geometries. This suggests that the spatial feature varies drastically for different array geometries, which significantly degrades binaural rendering performance. The MAC value for ICPD cannot even be calculated for two array configurations with different numbers of microphones because the dimensions of the ICPD vector are different. In contrast, the average of the off-diagonal MAC values in Tables III and IV have reached 63.22% and 97.70%, respectively. This indicates that SCORE as a spatial feature is robust across different geometries and numbers of microphones. The use of ERB also helps improve robustness to variations in array configuration. In summary, the ERB-scaled SCORE feature proves to be a reliable method for extracting spatial features across different array configurations, making it an advantageous choice for implementing array-based binaural rendering systems in real-world applications.

TABLE II
MAC CALCULATED USING THE ICPD FEATURE FOR
VARIOUS ARRAY CONFIGURATION

| MAC | G1 | G2 | G3 | G4 |
|-----|-----|-----|-----|-----|
| G1 | 1 | 0.040 | 0.344 | n/a |
| G2 | 0.040 | 1 | 0.024 | n/a |
| G3 | 0.344 | 0.024 | 1 | n/a |
| G4 | n/a | n/a | n/a | 1 |

TABLE III
MAC CALCULATED USING THE SCORE FEATURE FOR
VARIOUS ARRAY CONFIGURATION

| MAC | G1 | G2 | G3 | G4 |
|-----|-----|-----|-----|-----|
| G1 | 1 | 0.632 | 0.671 | 0.605 |
| G2 | 0.632 | 1 | 0.654 | 0.640 |
| G3 | 0.671 | 0.654 | 1 | 0.591 |
| G4 | 0.605 | 0.640 | 0.591 | 1 |

TABLE IV
MAC Calculated using The ERB-Scaled SCORE
Feature for Various Array Configuration

| MAC | G1 | G2 | G3 | G4 |
|-----|-----|-----|-----|-----|
| G1 | 1 | 0.977 | 0.981 | 0.974 |
| G2 | 0.977 | 1 | 0.979 | 0.980 |
| G3 | 0.981 | 0.979 | 1 | 0.971 |
| G4 | 0.974 | 0.980 | 0.971 | 1 |

### D. AT Performance Evaluated using Objective Indices

To quantify the performance of the binaural AT, we propose three objective indices. IPD and ILD are two important spatial cues for source location [29, 30]. In the mw-IPD and mw-ILD errors, the IPD and ILD errors are penalized more in the time-frequency bins with large magnitudes than in the small ones. That is,

$$\text{mw-IPDe} = \frac{1}{\sum_{l,f} \sigma(l,f)} \sum_{l,f} \sigma(l,f) \left| \angle \frac{Y_L(l,f)}{Y_R(l,f)} - \angle \frac{\hat{Y}_L(l,f)}{\hat{Y}_R(l,f)} \right|, \quad (15)$$

$$\text{mw-ILDe} = \frac{1}{\sum_{l,f} \sigma(l,f)} \sum_{l,f} \sigma(l,f) \left| 20\log_{10} \frac{|Y_L(l,f)|}{|Y_R(l,f)|} - 20\log_{10} \frac{|\hat{Y}_L(l,f)|}{|\hat{Y}_R(l,f)|} \right|,$$

where $\angle(\cdot)$ is the phase operator and $\sigma(l,f) = \left( |Y_L(l,f)| + |Y_R(l,f)| \right)/2$ is the average magnitude of left and right ears. We calculate mw-IPDe up to 1500 Hz because human listeners are unable to perceive IPD at frequencies above 1500 Hz [41]. In addition, mSI-SDR is defined to evaluate the rendering and enhancement performance as follows:

$$\text{mSI-SDR} = 20\log_{10} \frac{\|\eta s\|_2^2}{\|\hat{s} - \eta s\|_2^2}, \quad (16)$$

where $\mathbf{s} = \begin{bmatrix} \text{STFT}^{-1}(\mathbf{Y}_L) & \text{STFT}^{-1}(\mathbf{Y}_R) \end{bmatrix}$ and $\hat{\mathbf{s}} = \begin{bmatrix} \text{STFT}^{-1}(\hat{\mathbf{Y}}_L) & \text{iSTFT}^{-1}(\hat{\mathbf{Y}}_R) \end{bmatrix}$ represent the target and the estimated signal vectors formed by contacting the inverse STFT of the left and right ear signals and $\eta = \frac{\langle \hat{\mathbf{s}}, \mathbf{s} \rangle}{\langle \mathbf{s}, \mathbf{s} \rangle}$, where $\langle , \rangle$ denotes the inner product operator.

Tables V summarizes the signal enhancement and ambience preservation performance obtained using the proposed ACIS-BAT and the baselines (LBH, MIF, and MDF) for different array configurations. For DSP-based approaches, LBH is effective in reducing background noise and extracting the target speech for the subsequent HRTF filtering, while MIF preserves both the target speech and the ambient sounds. As a result, LBH is capable of superior rendering in scenarios that require signal enhancement. On the other hand, MIF performs well in preserving the ambience. It is worth noting that both DSP-based methods can easily adapt to different array configurations by adjusting the array signal model without the need for training. However, when the number of sources exceeds the number of microphones, the so-called "underdetermined" case, the mSI-SDR of the DSP-based methods is relatively low.

TABLE V
Comparisons of Signal Enhancement and Ambience
Preservation Performance in Terms of mw-IPDe, mw-ILDe, and
mSI-SDR for Different Array Configuration

| | Array Geometry | Method | mw-IPD | mw-ILD | mSI-SDR |
|---|---|---|---|---|---|
| Signal enhancement | G1 | LBH | 0.51 | 3.21 | -28.83 |
| | | MIF | 1.18 | 8.23 | -30.00 |
| | | MDF-SE | 0.76 | 3.17 | 2.13 |
| | | ACIS-BAT | **0.22** | **1.30** | **4.83** |
| | G2 | LBH | 0.55 | 3.21 | -28.60 |
| | | MIF | 1.37 | 9.44 | -30.31 |
| | | MDF-SE | 1.63 | 7.70 | -7.16 |
| | | ACIS-BAT | **0.39** | **2.89** | **3.09** |
| | G3 | LBH | 0.70 | 3.54 | -29.55 |
| | | MIF | 1.31 | 9.72 | -30.46 |
| | | MDF-SE | 1.67 | 8.77 | -10.55 |
| | | ACIS-BAT | **0.36** | **2.22** | **4.10** |
| | G4 | LBH | 0.56 | 3.32 | -28.16 |
| | | MIF | 1.34 | 8.99 | -31.13 |
| | | MDF-SE | N/A | N/A | N/A |
| | | ACIS-BAT | **0.28** | **1.96** | **4.28** |
| Ambience preservation | G1 | LBH | 0.78 | 13.67 | -28.64 |
| | | MIF | 0.80 | 5.99 | -28.48 |
| | | MDF-AP | 0.57 | 4.13 | 2.47 |
| | | ACIS-BAT | **0.38** | **2.21** | **4.39** |
| | G2 | LBH | 0.82 | 13.77 | -28.60 |
| | | MIF | 1.02 | 7.67 | -29.11 |
| | | MDF-AP | 0.92 | 5.10 | -3.56 |
| | | ACIS-BAT | **0.45** | **3.43** | **2.60** |
| | G3 | LBH | 0.79 | 13.46 | -29.92 |
| | | MIF | 0.99 | 7.94 | -28.88 |
| | | MDF-AP | 0.80 | 4.63 | -2.13 |
| | | ACIS-BAT | **0.43** | **2.81** | **3.74** |
| | G4 | LBH | 0.84 | 13.85 | -28.15 |
| | | MIF | 0.91 | 6.99 | -29.54 |
| | | MDF-AP | N/A | N/A | N/A |
| | | ACIS-BAT | **0.40** | **2.85** | **3.87** |

Two learning-based approaches, MDF and ACIS-BAT, are discussed next. MDF outperforms the DSP-based methods when the tested array geometry is drawn from the training set (G1). However, performance degrades significantly for the unseen array geometries (G2 and G3). In addition, MDF cannot be used for the G4 configuration because the input dimension of the MDFnet is fixed. In contrast to MDF, the ACIS-BRnet trained with the SCORE feature performs the best in signal enhancement and ambience preservation, even in the face of unseen array geometries and sensor counts. Note that MDF requires the training of two MDFnets (MDF-SE and MDF-AP) on the target signal with $\alpha$ set to 0 and 1 for signal enhancement and ambience preservation, respectively, while the ACIS-BAT can accommodate any scenario through the tunable parameter in the BRnet.

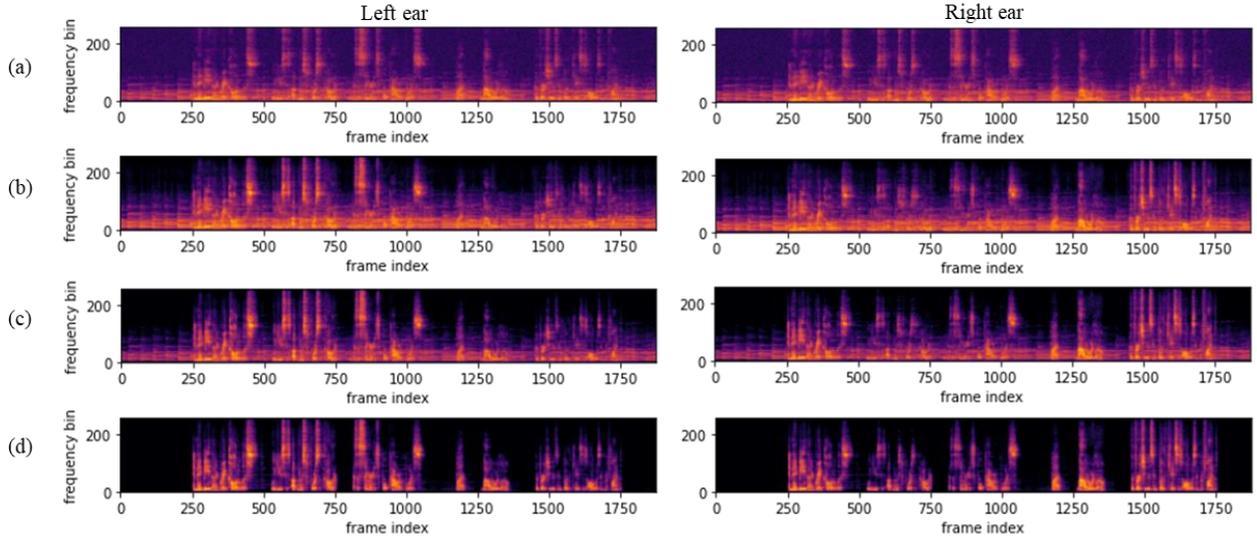

**Fig. 9.** Example spectrograms with 10 dB SAR and 20 dB SNR. (a) The noisy signal, the binaural outputs from BRnet with the parameter α set to (b) 1.0, (c) 0.5, and (d) 0.0.

### E. Scalability

To evaluate the scalability of the proposed ACIS-BAT system between signal enhancement and ambience preservation, we compare the mSI-SDR of the binaural signal outputs of the BRnet with the target signals in the signal enhancement mode (Eq. (2) with $\alpha = 0$) and in the ambience preservation mode (Eq. (2) with $\alpha = 0$), respectively. Figure 8 shows the mSI-SDR obtained using the BRnet for different $\alpha$. These results demonstrate the scalability of the proposed ACIS-BAT system for various application scenarios ranging from signal enhancement (teleconferencing) to ambience preservation (VR cinema). This is particularly advantageous compared to MDF, which requires training different models for different applications with no scalability during the testing phase. Figure 9 shows the results obtained with the ACIS-BAT system for the G1 configuration with T60 = 460 ms. It can be seen from the spectrograms that ACIS-BAT is able to balance the enhanced speech signals and the ambient noise according to the parameters specified.

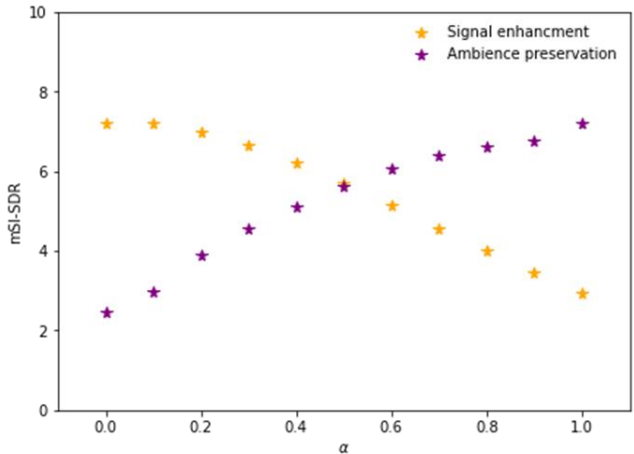

**Fig. 8.** The mSI-SDR of signal enhancement and ambience preservation plotted against the tunable parameter $\alpha$.

### F. Subjective Performance

A subjective listening test was performed using the Multiple Stimuli with Hidden Reference and Anchor (MUSHRA) procedure [42]. The subjective indices for ambience preservation include sense of direction, ambience preservation, dereverberation, sensor noise reduction, and overall quality on a scale of 1 to 5. On the other hand, the subjective indices for signal enhancement include sense of direction, background noise reduction, dereverberation, sensor noise reduction, and overall quality. To evaluate ambience preservation, the target signals in Eq. (2) are prepared by setting $\alpha = 0$, while for signal enhancement, $\alpha = 0$ is used. In addition, highpass filtered signals are used as the anchor. In total, 15 subjects of different genders, experts, and non-experts participated in the listening test. ANalysis Of VAriance (ANOVA) is employed to process the MUSHRA results with respect to ambience preservation (Fig. 10). In particular, the learning-based methods, MDF and ACIS-BAT, significantly outperform the DSP-based baselines, LBH and MIF, on all indices. Although LBH and MIF provide a satisfactory sense of direction, LBH falls short in maintaining a sense of ambience. On the other hand, MIF does not suppress sensor noise because it is not designed for that purpose. ACIS-BAT delivers a better sense of direction than MDF at the expense of sensor noise reduction. The variance of the overall performance of ACIS-BAT is small.

To investigate signal enhancement, we used the restaurant noise data selected from the Microsoft Scalable Noisy Speech Dataset (MS-SNSD) [43] to serve as an ambient sound different from the one used in the training phase. Figure 11 summarizes the MUSHRA results for the signal enhancement operation. MIF performs poorly in all aspects of background noise, reverberation, and sensor noise. In contrast, LBH outperforms MIF on all subjective metrics, but slightly underperforms MDF. Although MDF performs satisfactorily on most metrics, its performance on sense of direction is only acceptable. However, by using the proposed ACIS-BAT system, superior enhancement performance can be achieved if the parameter $\alpha$ is set to 0. In summary, DSP-based methods rely heavily on the knowledge of the array configurations and hence the signal model. While MDF can excel in both AT modes, dedicated MDF networks (MDF-AP and MDF-SE) need to be trained for signal enhancement or ambience preservation purposes. The proposed ACIS-BAT system can achieve optimal performance in a scalable manner to achieve a seamless balance between these two purposes.

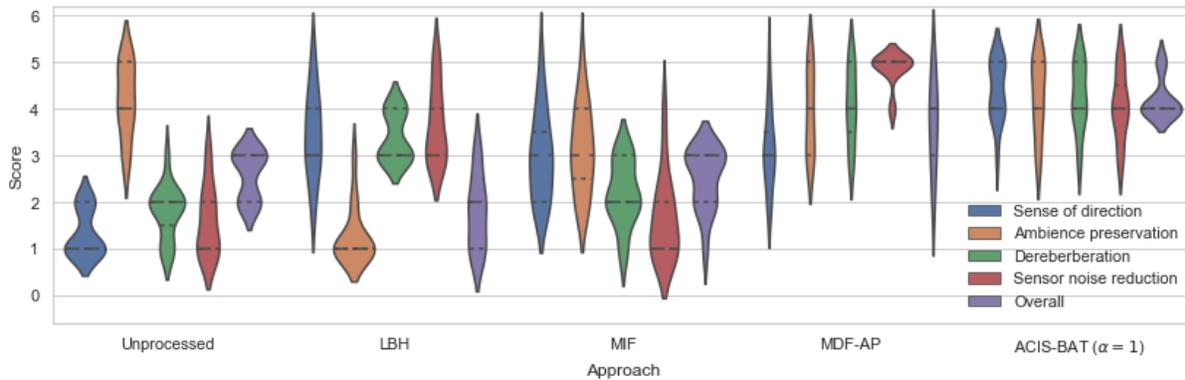

**Fig. 10.** The ANOVA output of the listening test for the ambience preservation mode.

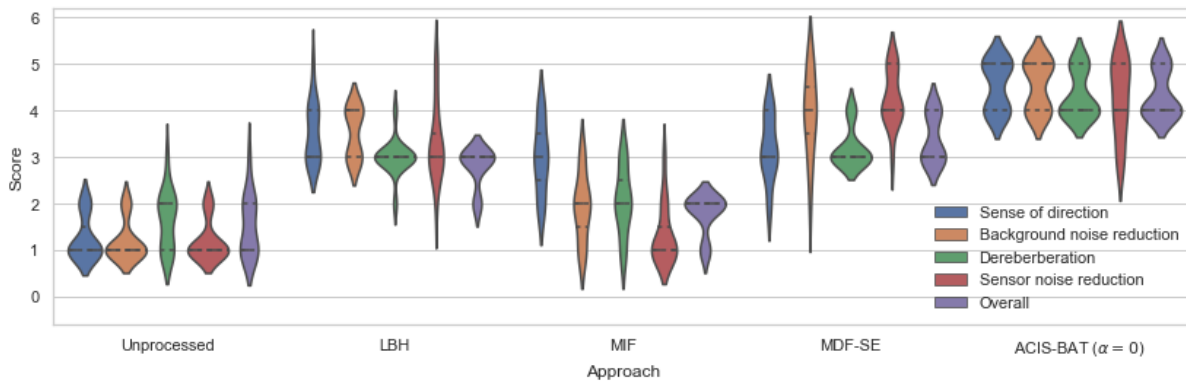

**Fig. 11.** The ANOVA output of the listening test for the signal enhancement mode.

## V. Conclusion

A learning-based ACIS-BAT system has been presented to convert microphone array signals into the binaural signals. It has been shown that the proposed SCORE is a spatial feature robust against unseen array configurations. As a core element, BRnet proves to be effective in signal enhancement and ambience preservation, requiring fewer parameters and less computational complexity than MDFnet. Furthermore, ACIS-BAT allows for scalability between signal enhancement and ambience preservation through the tunable parameter in BRnet. Future research will extend BAT to generalized global AT based on integrated microphone and loudspeaker arrays.

## References

[1] F. Z. Kaghat, A. Azough, M. Fakhour, and M. Meknassi, "A new audio augmented reality interaction and adaptation model for museum visits," *Comput. Electr. Eng.*, vol. 84, Jun. 2020, Art. no. 106606.

[2] A. C. Kern and W. Ellermeier, "Audio in VR: Effects of a soundscape and movement-triggered step sounds on presence," *Front. Robot. AI*, vol. 7, pp. 1–13, 2020.

[3] A. N. Nagele, V. Bauer, P. G. T. Healey, J. D. Reiss, H. Cooke, T. Cowlishaw, C. Baume, and C. Pike, "Interactive audio augmented reality in participatory performance," *Frontiers Virtual Reality*, vol. 1, pp. 1–14, 2021.

[4] J. Yang, A. Barde, and M. Billinghurst, "Audio augmented reality: A systematic review of technologies, applications, and future research directions," *J. Audio Eng. Soc.*, vol. 70, pp. 788–809, 2022.

[5] M. Zaunschirm, C. Schörkhuber, and R. Höldrich, "Binaural rendering of Ambisonic signals by head-related impulse response time alignment and a diffuseness constraint," *J. Acoust. Soc. Am.*, vol. 143, no. 6, pp. 3616–3627, 2018.

[6] R. Gupta, J. He, R. Ranjan, W. Gan, F. Klein, C. Schneiderwind, A. Neidhardt, K. Brandenburg, and V. Välimäki, "Augmented/mixed reality audio for hearables: Sensing, control, and rendering," *IEEE Signal Processing Magazine*, vol. 39, no. 3, pp. 63-89, 2022.

[7] B. Rafaely et al., "Spatial audio signal processing for binaural reproduction of recorded acoustic scenes—review and challenges," *Acta Acustica*, vol. 6, no. 47, 2022.

[8] M. R. Bai, Y. Chen, Y. Hsu, and T. Wu, "Robust binaural rendering with the time-domain underdetermined multichannel inverse prefilters," *J. Acoust. Soc. Am.*, vol. 146, no. 2, pp. 1302–1313, 2019.

[9] F. Lluís, P. Martínez-Nuevo, M. B. Møller, S. E. Shepstone, "Sound field reconstruction in rooms: Inpainting meets super-resolution," *J. Acoust. Soc. Am.*, vol. 148, no. 2, pp. 649–659, 2020.

[10] E. Fernandez-Grande, X. Karakonstantis, D. Caviedes-Nozal, and P. Gerstoft, "Generative models for sound field reconstruction," *J. Acoust. Soc. Am.*, vol. 153, no. 2, pp. 1179–1190, 2023.

[11] A. Wabnitz, N. Epain, A. McEwan, and C. Jin, "Upscaling Ambisonic sound scenes using compressed sensing techniques," *Proc. IEEE Workshop on Applications of Signal Processing to Audio and Acoustics (WASPAA)*, 2011, pp. 1–4.

[12] M. Kentgens, S. A. Hares and P. Jax, "On the Upscaling of Higher-Order Ambisonics Signals for Sound Field Translation," *Proc. European Signal Processing Conference (EUSIPCO)*, 2021, pp. 81–85.

[13] A. Berkhout, D. de Vries, and P. Vogel, "Acoustic control by wave field synthesis," *J. Acoust. Soc. Am.*, vol. 93, no. 5, pp. 2764–2778, 1993.

[14] J. Ahrens and S. Spors, "Wave field synthesis of a sound field described by spherical harmonics expansion coefficients," *J. Acoust. Soc. Am.*, vol. 131, no. 3, pp. 2190–2199, 2012.

[15] T. D. Abhayapala and D. B. Ward, "Theory and design of high order sound field microphones using spherical microphone array," *Proc. IEEE ICASSP*, 2002, pp. 1949–1952.

[16] D. L. Alon, J. Sheaffer, and B. Rafaely, "Robust plane-wave decomposition of spherical microphone array recordings for binaural sound reproduction," *J. Acoust. Soc. Amer.*, vol. 138, no. 3, pp. 1925–1926, 2015.

[17] W. Zhang, T. D. Abhayapala, R. A. Kennedy, and R. Duraiswami, "Insights into head-related transfer function: Spatial dimensionality and continuous representation," *J. Acoust. Soc. Am.*, vol. 127, no. 4, pp. 2347–2357, 2010.

[18] M. Jeffet, N. R. Shabtai, and B. Rafaely, "Theory and perceptual ecaluation of the binaural reproduction and beamforming trade-off in the generalized spherical array beamformer," *IEEE/ACM Trans, Audio, Speech, Lang. Proc.*, vol. 24, no. 4, pp. 708–718, 2016.

[19] Z. Ben-Hur, D. L. Alon, R. Mehra, and B. Rafaely, "Binaural reproduction based on bilateral ambisonics and ear-aligned hrtfs," *IEEE/ACM Trans, Audio, Speech, Lang. Proc.*, vol. 29, pp. 901–913, 2021.

[20] L. Birnie, Z. Ben-Hur, V. Tourbabin, T. Abhayapala, and P. Samarasinghe, "Bilateral-Ambisonic Reproduction by Soundfield Translation," *Proc. Intl. Workshop Acoustic. Signal Enhancement (IWAENC)*, 2022, pp. 1–5.

[21] C. Borrelli, A. Canclini, F. Antonacci, A. Sarti, and S. Tubaro, "A denoising methodology for higher order ambisonics recordings," *Proc. Intl. Workshop Acoustic. Signal Enhancement (IWAENC)*, 2018, pp. 451–455.

[22] M. Lugasi and B. Rafaely, "Speech enhancement using masking for binaural reproduction of ambisonics signals," *IEEE/ACM Trans, Audio, Speech, Lang. Proc.*, vol. 28, pp. 1767–1777, 2020.

[23] A. Herzog and E. A. Habets, "Direction and reverberation preserving noise reduction of Ambisonics signals," *IEEE/ACM Trans, Audio, Speech, Lang. Proc.*, vol. 28, pp. 2461–2475, 2020.

[24] H. Beit-On, M. Lugasi, L. Madmoni, A. Kumar, J. Donley, V. Tourbabin, and B. Rafaely, "Audio signal processing for telepresence based on wearable array in noisy and dynamic scenes," *Proc. IEEE ICASSP*, 2022, pp. 8797–8801.

[25] I. Ifergan and B. Rafaely, "On the selection of the number of beamformers in beamforming-based binaural reproduction," *EURASIP Journal on Audio, Speech, and Music Proc.*, 6, 2022.

[26] Y. Hsu, C. Ma, and M. R. Bai, "Model-matching principle applied to the design of an array-based all-neural binaural rendering system for audio telepresence," *Proc. IEEE ICASSP*, 2023, pp. 1–5.

[27] H. Schroter, A. N. Escalante-B., T. Rosenkranz, and A. Maier, "Deepfilternet: A low complexity speech enhancement framework for full-band audio based on deep filtering," *Proc. IEEE ICASSP*, 2022, pp. 7407–7411.

[28] E. Perez, F. Strub, H. De Vries, V. Dumoulin, and A. Courville, "Film: Visual reasoning with a general conditioning layer," *Proc. 32nd AAAI Conf. Artif. Intell.*, 2018, pp. 3942–3951.

[29] B. G. Shinn-Cunningham, S. G. Santarelli, and N. Kopco, "Tori of confusion: Binaural cues for sources within reach of a listener," *J. Acoust. Soc. Amer.*, vol. 107, no. 3, pp. 1627-1636, 2000.

[30] J. Braasch, "Modelling of binaural hearing," in *Communication Acoustics*, New York: Springer, pp. 75-108, 2005.

[31] K. Scharnhorst, "Angles in complex vector spaces", *Acta Applicandae Mathematicae*, vol. 69, no. 1, pp. 95-103, 2001.

[32] B. C. F. Moore. *An introduction to the psychology of hearing*. Brill, 2012.

[33] A. Ephrat, I. Mosseri, O. Lang, T. Dekel, K. Wilson, A. Hassidim, W. T. Freeman, and M. Rubinstein, "Looking to listen at the cocktail party: A speaker-independent audio-visual model for speech separation," *ACM Trans. Graph.*, vol. 37, no. 4, 2018.

[34] V. Panayotov, G. Chen, D. Povey, and S. Khudanpur, "Librispeech: an ASR corpus based on public domain audio books," *Proc. IEEE ICASSP*, 2015, pp. 5206–5210.

[35] C. Armstrong, L. Thresh, D. Murphy, and G. Kearney, "A perceptual evaluation of individual and non-individual HRTFs: a case study of the SADIE II database," *Applied Sciences*, vol. 8, no. 11, 2018.

[36] E. Lehmann and A. Johansson, "Prediction of energy decay in room impulse responses simulated with an image-source model," *J. Acoust. Soc. Amer.*, vol. 124, no. 1, pp. 269–277, Jul. 2008.

[37] C. Armstrong, L. Thresh, D. Murphy, and G. Kearney, "A perceptual evaluation of individual and non-individual HRTFs: a case study of the SADIE II database," *Applied Sciences*, vol. 8, no. 11, 2018.

[38] S. Braun, H. Gamper, C. K. A. Reddy, and I. Tashev, "Towards efficient models for real-time deep noise suppression," *Proc. IEEE ICASSP*, 2021, pp. 656–660.

[39] J. H. DiBiase, H. F. Silverman, and M. S. Brandstein, "Robust localization in reverberant rooms" in Microphone Arrays, Berlin, Germany:Springer, pp. 157–1880, 2001.

[40] H. L. V. Trees, Optimum array processing, detection, estimation, and modulation theory, Part IV, New York: Wiley, 2002.

[41] A. Brughera, L. Dunai, and W. M. Hartmann, "Human interaural time difference thresholds for sine tones: the high-frequency limit," *J. Acoust. Soc. Am.*, vol. 133, pp. 2839–2855, 2013.

[42] ITU-R Recommendation, "Method for the subjective assessment of intermediate sound quality (MUSHRA)," *International Telecommunications Union*, BS. 1534-1, Geneva, Switzerland, 2001.

[43] C. K. Reddy, E. Beyrami, J. Pool, R. Cutler, S. Srinivasan, and J. Gehrke, "A scalable noisy speech dataset and online subjective test tramewoek," *Proc. Interspeech*, 2019, pp. 1816–1820.